\documentclass[aps,12pt,nofootinbib,showpacs,showkeys,preprintnumbers,amsmath,amssymb]{revtex4-1}		

\usepackage{amsmath,amsfonts,amssymb,amsthm,bm,bbm,bbold,braket,color,dsfont,fancybox,srcltx,subcaption,graphicx,ucs,yfonts,cancel,xfrac,verbatim,xcolor,slashed}

\usepackage[colorlinks=true,
urlcolor=darkraspberry,
linkcolor=darkraspberry,
citecolor=darkraspberry,
linktocpage=true,
pdfproducer=medialab,
pdfa=true,
anchorcolor=blue]{hyperref}%
\definecolor{cambridgeblue}{rgb}{0.64, 0.76, 0.68}
\definecolor{darkraspberry}{rgb}{0.53, 0.15, 0.34}

\usepackage{float}

\usepackage[section]{placeins}

\usepackage{lineno}

\usepackage{xcolor}
\usepackage{orcidlink}   

\definecolor{GCcolor}{RGB}{88,24,124}    
\definecolor{ZSWcolor}{RGB}{0,102,204}   
\definecolor{BDScolor}{RGB}{0,153,76}    
\definecolor{PAcolor}{RGB}{204,102,0}    
\definecolor{YUcolor}{RGB}{180,0,0}      

\usepackage[normalem]{ulem}

\begin{document}

\title{Can LLP detectors probe the reheating temperature? A case study of vector dark matter}

\author{Paulo Areyuna C.$^{1,2}$\,\orcidlink{0000-0003-0061-4961}}
\email{pareyunac@estudiante.uc.cl}

\author{Giovanna Cottin$^{1,2}$\,\orcidlink{0000-0002-5308-5808}}
\email{gfcottin@uc.cl}

\author{Bastián Díaz Sáez$^{2,1}$\,\orcidlink{0000-0002-7386-8113}}
\email{bastian.diaz@uc.cl}

\author{Zeren Simon Wang$^{3}$\,\orcidlink{0000-0002-1483-6314}}
\email{wzs@hfut.edu.cn}

\author{Yu Zhang$^{3}$\,\orcidlink{0000-0001-9415-8252}}
\email{dayu@hfut.edu.cn}

\affiliation{$^1$Instituto de Física, Pontificia Universidad Católica de Chile, Avenida Vicuña Mackenna 4860, Santiago, Chile}

\affiliation{$^2$Millennium Institute for Subatomic Physics at the High Energy Frontier (SAPHIR), \\ Fern\'andez Concha 700, Santiago, Chile}

\affiliation{$^3$School of Physics, Hefei University of Technology, Hefei 230601, China}

\begin{abstract}
We study an extension of the singlet-scalar Higgs portal featuring a dark vector $V_\mu$ and a real scalar $\phi$. The vector is a dark matter (DM) candidate, while $\phi$ is long-lived and decays via higher-dimensional operators. We explore the DM production via freeze-in at low and high reheating temperatures. At colliders, the decay $\phi\to Z+V$ yields distinctive long-lived particle (LLP) signatures. We explore the interplay between cosmological constraints and LLP searches at the LHC and FCC-hh, showing that far detectors can probe otherwise inaccessible parameter space and place novel bounds on the reheating temperature.
\end{abstract}


\maketitle
\tableofcontents

\section{Introduction}
The nature of dark matter (DM) remains one of the most compelling open problems in modern particle physics.
Minimal extensions of the Standard Model (SM), in which the observed relic abundance is explained by the addition of a single new field, are increasingly constrained by a wide range of experimental searches, particularly direct detection experiments.
This has motivated the exploration of simple yet non-minimal dark sector frameworks, which can more readily evade current bounds while offering a richer and potentially testable phenomenology, especially at colliders and dedicated far detector experiments.

The freeze-in mechanism~\cite{Hall:2009bx} provides a well-motivated way to evade these constraints, as the DM relic abundance is generated through extremely feeble interactions with the SM.
However, the same small couplings that ensure consistency with existing bounds also tend to suppress observable signals, making these scenarios notoriously difficult to probe experimentally

In recent years, renewed interest in freeze-in has emerged in the context of low reheating temperatures~\cite{Frangipane:2021rtf, Cosme:2023xpa, Silva-Malpartida:2023yks, Boddy:2024vgt} (see also ref.~\cite{Belanger:2018sti}), opening new avenues for testing such models~\cite{Barman:2024nhr,Barman:2024tjt,Borah:2025ema}.
In particular, in scenarios with two new fields, this framework naturally predicts the coexistence of DM and a heavier, unstable, and long-lived mediator.
This feature provides distinctive experimental signatures~\cite{Alimena:2019zri,Lee:2018pag}, rendering these models especially well suited for searches at colliders and far detector experiments. 

In this work, we study the DM  and collider phenomenology of a simple extension of the SM based on the introduction of a new $U(1)_V$ gauge symmetry, which leads to the presence of a massive vector boson $V$, plus a real scalar singlet $\phi$, with both new fields odd under a new $Z_2$ symmetry.
This scenario with  $\phi B_{\mu\nu}\tilde{V}^{\mu\nu}$  is quite similar to the so called \textit{dark-axion portal}~\cite{Kaneta:2016wvf}, where $B^{\mu\nu}$ and $V^{\mu\nu}$ are the corresponding field strengths of the $U(1)_Y$ and $U(1)_V$ gauge groups, respectively.
The dark-axion portal has attracted the attention in the last decade from collider and reactor physics~\cite{deNiverville:2018hrc, Deniverville:2020rbv, Jodlowski:2023yne, Jodlowski:2024ayf}, cosmology and astrophysics~\cite{Hook:2021ous, Hong:2023fcy}, and dark matter~\cite{Kaneta:2017wfh, Arias:2020tzl, Gutierrez:2021gol, Broadberry:2024pkv, DiazSaez:2024dzx, Arias:2025nub, Arias:2025tvd}. Besides the dark-axion portal, this setup presents a Higgs-portal $\sim \phi^2|H|^2$.

As we focus on the scenario in which $V$ is the lightest dark-sector particle and constitutes a DM candidate, $\phi$ plays the role of an unstable mediator.
In particular, we focus on the case where $\phi$ is a long-lived particle relevant to be searched for at colliders and proposed far detector experiments.
The decay channel that has been studied for detection at colliders~\cite{deNiverville:2018hrc, Deniverville:2020rbv, Jodlowski:2023yne, Jodlowski:2024ayf, DiazSaez:2024dzx, Arias:2025nub} is
\begin{eqnarray}\label{channel1}
\phi  \to  \gamma + V,
\end{eqnarray}
which results in striking signatures involving non-pointing photons.
However, the model also predicts the decay channel
\begin{eqnarray}\label{channel2}
\phi  \to  Z + V,
\end{eqnarray}
with an on-shell $Z$-boson, which has not been explored previously in detail.\footnote{In freeze-out scenarios, in particular in the co-scattering regime, the mass splitting between the two dark states is typically too small to allow an on-shell $Z$ for masses in the ballpark of $\mathcal{O}$(GeV-TeV), while in previously studied freeze-in scenarios the mediator mass was restricted to values below $m_Z$ (see e.g.~ref.~\cite{Arias:2025tvd}).}
We fill this gap by investigating the regime $m_\phi > m_Z$, allowing us to fully exploit the phenomenology associated with the decay channel in equation~\eqref{channel2}, and to assess its discovery prospects at the LHC and far detector experiments.

Long-lived particle (LLP) signatures are particularly timely in view of the upcoming experimental programs dedicated to LLP searches, including proposed and ongoing experiments as MATHUSLA~\cite{Chou:2016lxi,Curtin:2018mvb,MATHUSLA:2018bqv,MATHUSLA:2020uve,MATHUSLA:2025zyt}, FASER~\cite{Feng:2017uoz,FASER:2018eoc,FASER:2022hcn}, CODEX-b~\cite{Gligorov:2017nwh,Aielli:2019ivi}, and ANUBIS~\cite{Bauer:2019vqk,ANUBIS:2025sgg}.
These detectors offer unprecedented sensitivities to weakly coupled new states with macroscopic decay lengths, providing a powerful and complementary probe to conventional collider searches.
Moreover, the Future Circular Collider (FCC)~\cite{FCC:2018vvp} in its hadron-hadron mode FCC-hh~\cite{Benedikt:2022kan} and its associated proposed far detectors~\cite{Bhattacherjee:2021rml,Bhattacherjee:2023plj} are expected to have even more promising search prospects, particularly for LLPs at the $\mathcal{O}$(GeV) scale.

The paper is structured as follows.
In Sec.~\ref{sec:model} we introduce our model setup, including the parameter ranges of our interest and the $\phi$-lifetime computation.
We then discuss in Sec.~\ref{sec:dm} the relevant DM constraints.
We present the collider phenomenology of our model scenario in Sec.~\ref{sec:collider}, studying the prospects at both main and far detectors at the LHC and the FCC-hh.
The numerical results are then displayed and discussed in Sec.~\ref{sec:results}, before we conclude in Sec.~\ref{sec:conclusions}. In Appendix~\ref{App} we present a simple analysis considering extra radiation during the epoch of Big-Bang Nucleosynthesis (BBN).

\section{The model}\label{sec:model}
We introduce a dark sector consisting of a massive vector boson $V_\mu$ from the $U(1)_V$ new gauge group, and a real  scalar singlet $\phi$.
Both fields are assumed to be odd under a new $Z_2$ symmetry, whereas all SM fields are even.
In this scenario, the vector field acquires mass via the Stueckelberg mechanism \cite{Ruegg:2003ps}, making it a viable DM candidate without breaking gauge invariance.
The interaction between the dark sector and the SM is mediated through a Higgs-portal coupling and a dimension-5 operator coupling the scalar singlet to the hypercharge field strength tensor $B_{\alpha\beta}$.
\footnote{In principle, the Stueckelberg mechanism can be used to generate a term of the form $\phi^2 V_\mu V^\mu$ without breaking gauge invariance. However, this is only posible with dimension six operators, and our EFT is restricted to dimension 5. Gauge invariance also forbids other dim5 operators such as $\frac{1}{\Lambda} \phi V_\mu (H^\dagger i\stackrel{\leftrightarrow}{D^\mu}  H)$.
}. The Lagrangian of the model is given by~\cite{DiazSaez:2024dzx}:
\begin{equation}
\begin{split}
    \mathcal{L} \; =& \; \mathcal{L}_{\text{SM}} 
    - \frac{1}{4} V_{\mu\nu} V^{\mu\nu} 
    + \frac{1}{2} m_V^2 V_\mu V^\mu 
    + \frac{1}{2} \partial_\mu \phi \, \partial^\mu \phi 
    - \frac{1}{2} m_\phi^2 \phi^2 \\
    & \; - \lambda_\phi \, \phi^4 
    - \lambda_{HS} \, \phi^2 |H|^2 
    + \frac{c_5}{\Lambda} \, \phi \, V_{\mu\nu} \, \hat{B}^{\mu\nu},
\end{split}
\label{modelLag}
\end{equation}
with 
\[
    \hat{B}^{\mu\nu} \equiv \frac{1}{2} \epsilon^{\mu\nu\alpha\beta} B_{\alpha\beta},
\]
where $H$ is the SM Higgs doublet, and $V_{\mu\nu} = \partial_\mu V_\nu - \partial_\nu V_\mu$.
The parameters $m_V$ and $m_\phi$ denote the masses of the vector boson and scalar singlet, respectively.
The couplings $\lambda_\phi$ and $\lambda_{HS}$ represent the self-interaction of the scalar singlet and its interaction with the Higgs field, respectively.
The scale $\Lambda$ characterizes the suppression of the non-renormalizable operator, and $c_5$ is a dimensionless Wilson coefficient.
To establish bounds on the cutoff scale, we will fix $c_5=1$ from now on.

The model can be regarded as an extension of the well-known singlet-scalar Higgs-portal scenario~\cite{OConnell:2006rsp,Wells:2008xg,Bird:2004ts,Pospelov:2007mp,Krnjaic:2015mbs,Boiarska:2019jym}.
As $m_\phi > m_{ V }$, the inclusion of non-renormalizable interactions allows the singlet scalar $\phi$ to decay.
As a consequence, the vector state $V_\mu$ remains as the only stable DM candidate.

In this work, we focus on the following benchmark ranges for the parameters:
\begin{align*}
    m_{V} &\in [0.01,\, 50]~\text{GeV}, \\
    m_{\phi} &\in \left[M_Z,\, 500\right]~\text{GeV}, \\
    \lambda_{HS} &\in [10^{-2},\, 3], \\
    \Lambda &\in [10^{8},\, 10^{12}]~\text{GeV}. 
\end{align*}
This parameter space is complementary to the one explored in a previous study~\cite{Arias:2025tvd}, which focused on lighter dark-state masses.

\subsection{Decays of the long-lived mediator $\phi$}

The decay width of $\phi$ has two main contributions:
\begin{eqnarray}
    \Gamma_{\phi} = \Gamma_{\phi \to V \gamma} + \Gamma_{\phi \to V Z},
    \label{eqn:Gamma_phi}
\end{eqnarray}
where 
\begin{eqnarray}
\Gamma(\phi \to \gamma V)
=
\frac{\cos^2\theta_W}{8\pi \Lambda^2}\, m_\phi^3
\left(1 - \frac{m_{V}^2}{m_\phi^2}\right)^3,
\label{eqn:Gamma_phi2gammaV}
\end{eqnarray}
with $\theta_W$ labeling the weak mixing angle.
The partial decay width of $\phi\rightarrow ZV$ is given by
\begin{eqnarray}
\Gamma(\phi \to Z V)
=
\frac{\sin^2\theta_W}{8\pi \Lambda^2}\,
m_\phi^3
\left[
\left(1-\frac{(m_Z+m_{V})^2}{m_\phi^2}\right)
\left(1-\frac{(m_Z-m_{V})^2}{m_\phi^2}\right)
\right]^{3/2},
\label{eqn:Gamma_phi2ZV}
\end{eqnarray}
where $m_Z$ is the $Z$-boson mass.
Equation~\eqref{eqn:Gamma_phi2ZV}, in the limit $m_V \ll m_\phi$, reduces to
\begin{eqnarray}
\Gamma(\phi \to Z V)
\simeq
\frac{\sin^2\theta_W}{8\pi \Lambda^2}
m_\phi^3
\left(1-\frac{m_Z^2}{m_\phi^2}\right)^3.
\end{eqnarray}

The total decay width of $\phi$ is dominated by $\Gamma(\phi\to\gamma V)$, and thus the proper decay length of $\phi$ can be approximated as
\begin{eqnarray}
c\tau_\phi \simeq
0.64\,{\rm m}\,
\left(\frac{\Lambda}{10^{10}\,{\rm GeV}}\right)^2
\left(\frac{100\,{\rm GeV}}{m_\phi}\right)^3
\left(1-\frac{m_V^2}{m_\phi^2}\right)^{-3}.
\label{ctau}
\end{eqnarray}

\begin{figure}[t]
    \centering
    \includegraphics[width=0.495\linewidth]{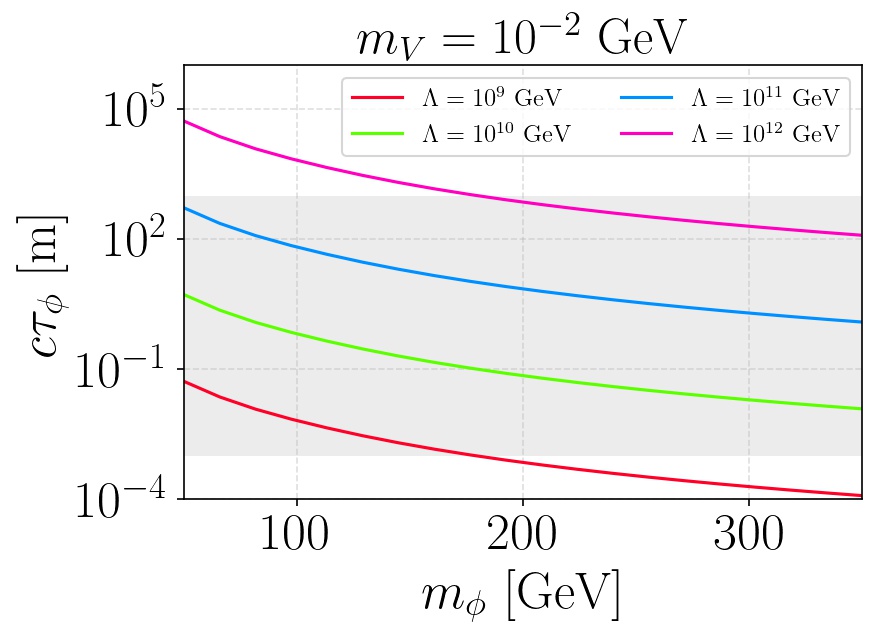}
    \caption{ $c\tau_\phi$ as function of $m_\phi$ for different values of $\Lambda$. The gray region for $c\tau_\phi\in[10^{-3},10^3]$~m highlights the LLP target region. 
    }
    \label{fig:ctau}
\end{figure}

In figure~\ref{fig:ctau} we show a plot of $c\tau_\phi$ as function of $m_\phi$. We fix $m_V=10^{-2}$~GeV, and display curves for various values of $\Lambda$. 
\begin{figure}
    \centering
    \includegraphics[width=0.45\linewidth]{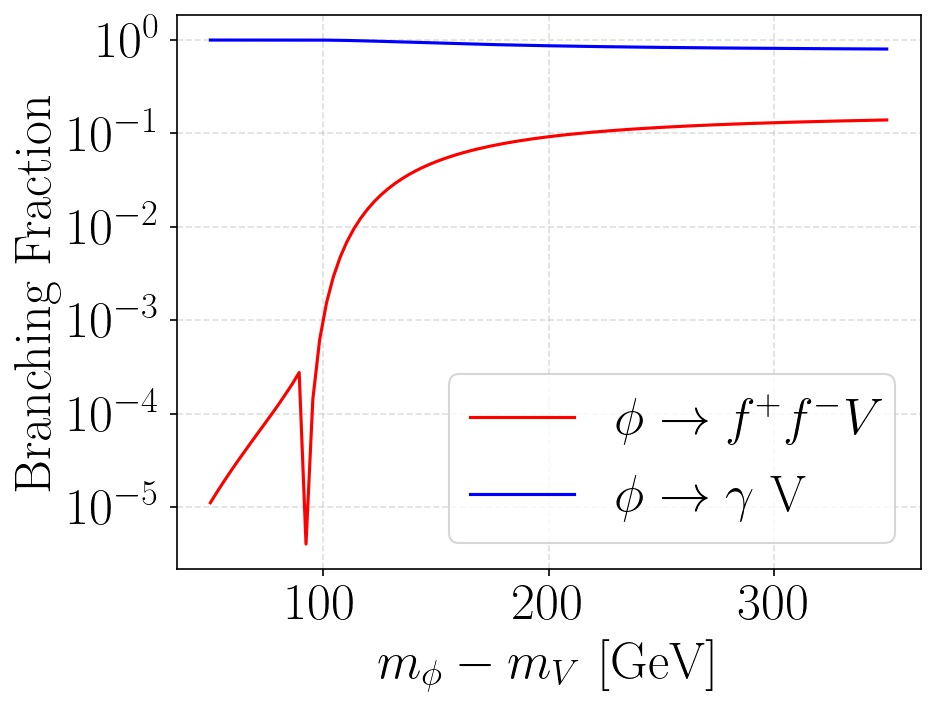}
    \caption{Branching fractions of the scalar $\phi$ as functions of $m_\phi - m_V$. }
    \label{branching}
\end{figure}
While the important decays for DM phenomenology are $\phi \to \gamma  V$ and $\phi \to ZV$, the study of collider signatures requires to consider the decay of the scalar to charged particles: $\phi \to f^+ f^- V$. For the kinematical regime where $m_\phi-m_V>m_Z$, this branching fraction can be estimated as

\begin{equation}
    \text{Br}(\phi \to f^+ f^-V)=  \text{Br}(\phi \to ZV) \times \text{Br}(Z \to f^+f^-).
\end{equation}
In this kinematic regime, Higgs observables remain SM-like: the decay $h\to \phi\phi$ is kinematically forbidden, and the $Z_2$
 symmetry prevents Higgs–scalar mixing. This allows us to safely explore this region of parameter space.
 We used {\texttt{MadGraph5\_aMC@NLO}}~\cite{Alwall:2011uj,Alwall:2014hca} version 3.6.2 to compute these branching fractions, as depicted in figure \ref{branching}. Although the decay channel of $\phi \to f^+f^-V$ is subdominant, this signature is potentially sizable and possible to detect at both main and far detectors, complementing existing searches for non-pointing photons which make use of calorimeter activity to reconstruct the displaced signal (see e.g.~ref.~\cite{Arias:2025tvd}).

\section{Dark matter, constraints, and cosmological tests}\label{sec:dm}
\subsection{Boltzmann equations}

In this section, we discuss the production mechanisms of the dark photon $V$ in the early Universe, which include freeze-in (FI) and Super-WIMP (SW) contributions.
We consider $\lambda_{HS}$ sizable enough to thermalize $\phi$, and at some temperature it freezes out.
The dark photon $V$ is produced via freeze-in from the thermal bath and from the late decay of the frozen-out $\phi$ particles.
As we consider that the dark photon $V$ has small interactions with the thermal bath, we neglect back-reactions. In this way, the Boltzmann equations for the dark photon and the scalar singlet are given by,
\begin{eqnarray} \label{beq1}
  \label{eq:beq1}\frac{dY_{ V }}{dT}  &=& -\frac{1}{\overline{H} T s}\left(\sum_f C_{f\bar{f}\rightarrow \phi V } + C_{W^+ W^-\rightarrow \phi V } + C_{\phi\rightarrow\gamma V (Z)}\right), \\ \label{beq2}
  \frac{dY_{\phi}}{dT}  &=& -\frac{1}{\overline{H} T s}\left(\sum_{X = f,b} C_{X\bar{X}\leftrightarrow \phi\phi}  - C_{\phi\rightarrow\gamma V (Z)}\right),  
\end{eqnarray}
where $H/\overline{H} \equiv 1 + \frac{T}{3}\frac{d\log g_{*s}}{dT}$, the Hubble rate is given by $H = \sqrt{\frac{\pi^2g_*}{90}}\frac{T^2}{M_P}$ with $g_*(T)$ denoting the effective number of relativistic degrees of freedom at temperature $T$, and $M_P = 2.4\times 10^{18}$~GeV is the reduced Planck mass.
The $C$ factors correspond to the integrated collision terms for the different processes contributing to the production and annihilation of $V$ and $\phi$.
The explicit expressions for these terms can be found in ref.~\cite{Arias:2025tvd}.
We have neglected the terms involving processes at $\mathcal{O}(g_D^4)$ such as $\gamma\gamma\rightarrow VV$ for being subleading at small $g_D$.\footnote{In the notation of ref.~\cite{Arias:2025tvd}, $g_D$ is a dimensionful parameter as $g_{D}=1/\Lambda$.}
Notice that unlike in ref.~\cite{Arias:2025tvd}, here the $Z$-boson can also be produced in the decay of $\phi$ because we study larger masses of $\phi$. 
 
In order to obtain the relic abundance of $V$, we make use of the freeze-in module of \texttt{micrOMEGAs}~\cite{Belanger:2018ccd} version 6.2.3.
Before  presenting the computation, in the following we  provide a discussion that significantly simplifies the problem.

\subsection{Thermalization of $\phi$ and Super-WIMP contribution}
\begin{figure}[t]
    \centering
    \includegraphics[width=\linewidth]{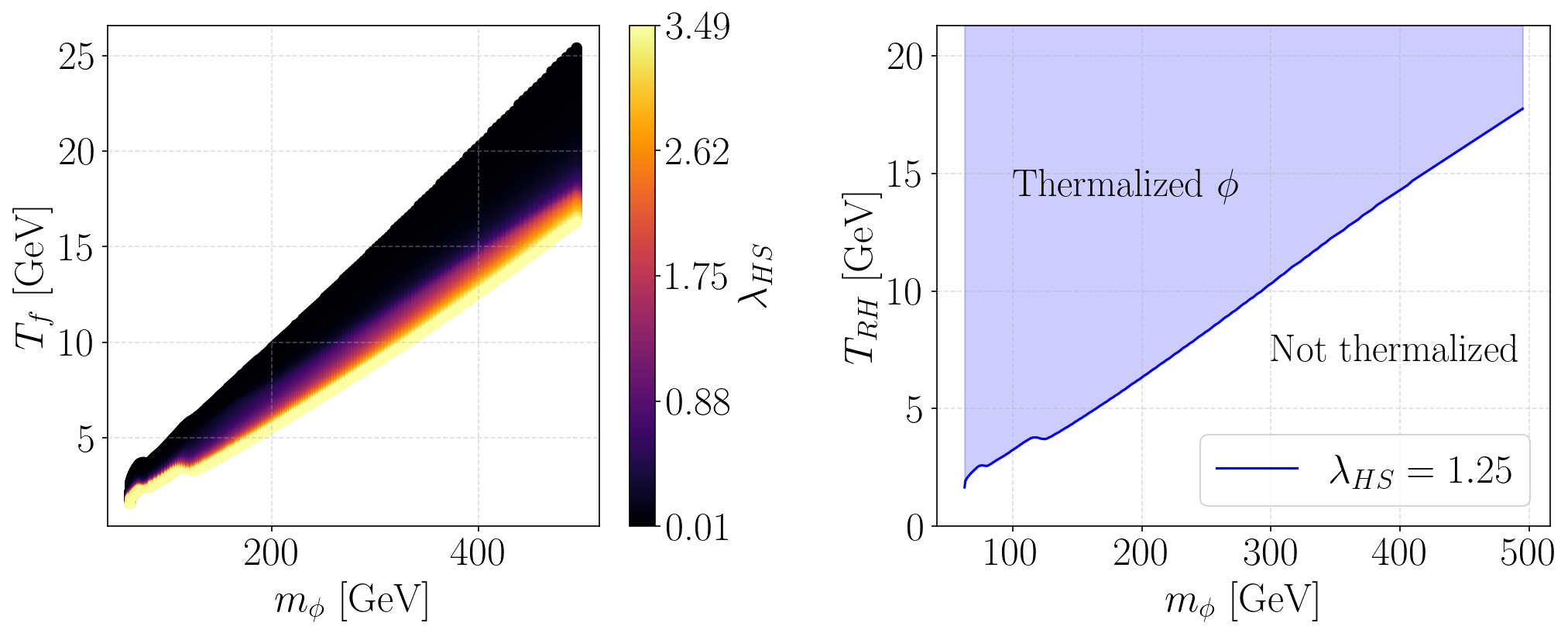}
    \caption{Left: Decoupling temperature of $\phi$ as a function of $m_\phi$ and $\lambda_{HS}$.
Right: Region of reheating temperatures for which $\phi$ reaches thermal equilibrium, shown for a fixed value of $\lambda_{HS}$. Larger values of $\lambda_{HS}$ enlarge the thermalization region, while smaller couplings reduce it.}
    \label{fig:thermalization}
\end{figure}

The scalar singlet $\phi$ achieves chemical equilibrium when its production and annihilation rates balance the expansion of the Universe, expressed by
\begin{equation}
\Gamma \equiv \langle \sigma v \rangle n_{\phi}^{\rm eq} \gtrsim H,
\end{equation}
where $\langle \sigma v \rangle$ is the thermally averaged cross section and $n_{\phi}^{\rm eq}$ is the equilibrium number density.
The interaction rate $\Gamma$ depends on the Higgs-portal coupling $\lambda_{HS}$, the mass of the scalar singlet $m_\phi$, and the temperature of the thermal bath $T$.
The Hubble rate $H$ is determined by a radiation dominated universe.
The interaction rate $\Gamma$ receives contributions from various annihilation and production channels involving SM particles:
\begin{eqnarray}
    \phi\phi \leftrightarrow f\bar{f}, W^+W^-, ZZ, hh, 
\end{eqnarray}
where $f$ denotes SM fermions. When $\Gamma$ drops below $H$, $\phi$ decouples from the thermal plasma at a temperature $T_f$, which is determined uniquely by $m_\phi$ and $\lambda_{HS}$, as a typical WIMP.
This is shown in the left panel of figure~\ref{fig:thermalization} as a color map.
On the other hand, we require $T_{\rm th} \leq T_{\rm RH}$ in order for $\phi$ to thermalize with the plasma. Here, $T_{\rm RH}$ denotes the maximum temperature of the thermal bath, and we work under the instantaneous reheating approximation. This approximation neglects the detailed dynamics of the reheating phase, which can be highly model dependent, and therefore provides a conservative and partial description of the cosmological history.

By identifying the minimum reheating temperature as $T_f$, and getting $T_f$ from \texttt{micrOMEGAs}, we can determine the corresponding values of $T_{\rm RH}$ to thermalize $\phi$ as a function of $(m_\phi, \lambda_{HS})$. This is illustrated in the right panel of figure~\ref{fig:thermalization}, where the blue region corresponds to the parameter space in which $\phi$ reaches thermal equilibrium for $\lambda_{HS} = 1.25$. The white region indicates not thermalization, as the interaction rates become kinematically suppressed. Increasing (decreasing) $\lambda_{HS}$ shifts the corresponding curve downward (upward).

\begin{figure}[t]
    \centering
    \includegraphics[width=0.55\linewidth]{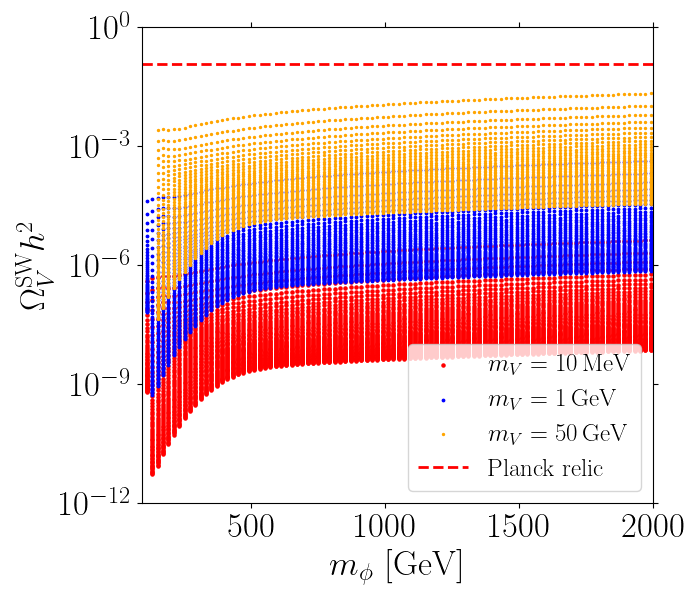}
    \caption{Super-WIMP contribution to the relic abundance for different dark photon masses and $\lambda_{HS} > 0.1$. All the points fulfill the condition $m_\phi > m_Z + m_{V}.$}
    \label{fig:sw}
\end{figure}

On the other hand, once $\phi$ has chemically decoupled from the SM plasma, at some lower temperature (the so-called \textit{decay temperature} $T_D$), it decays into $V$ and SM particles. This injection of DM from an out-of-equilibrium unstable state is known as the \textit{SuperWIMP} contribution~\cite{Feng:2003xh}, whose relic abundance is given by:
 \begin{eqnarray}
     \Omega_{ V }^{\text{SW}} h^2 = \frac{m_{ V }}{m_\phi} \, \Omega_{\phi} h^2,
 \end{eqnarray}
where $\Omega_{\phi} h^2$ is the relic abundance of the scalar singlet at freeze-out, computed using \texttt{micrOMEGAs}. In the parameter region we consider, the SW contribution is always subleading with respect to FI.
In figure~\ref{fig:sw}, we show its contribution to the relic abundance as a function of $m_\phi$ for several benchmark values of the dark photon mass: $m_V = 10$~MeV (red points), $m_V = 1$~GeV (blue points), and $m_V = 50$~GeV (orange points), assuming $\lambda_{HS} > 0.1$.
The dashed red line indicates the observed dark-matter relic abundance from Planck~\cite{Planck:2018vyg}.

As can be seen, the SW contribution is insufficient to account for the total relic abundance, or even for a $\sim 10$\% fraction of it.
It can become relevant only for $\lambda_{HS} < 0.1$ or a much heavier $\phi$, corresponding to parameter regions outside of our target.
Therefore, in the following, we neglect the Super-WIMP contribution and focus on the freeze-in production of the dark photon (for a related setup, see ref.~\cite{Garny:2018ali}).

\subsection{Freeze-in contribution}

Having established that the SW contribution is negligible in the parameter region under consideration, we study the FI production of dark photons.
The dominant processes arise from $\phi$ decays and fermion annihilation.
The resulting relic abundance of the dark photon, $\Omega_{V} h^2$, is given by
\begin{equation}
    \Omega_{V}^{\text{FI}} h^2 \simeq 2.82 \times 10^8 \, m_{V} \, Y_{V}(T_0),
\end{equation}
where $Y_{V}(T_0)$ denotes the present-day yield, obtained numerically using \texttt{micrOMEGAs}, and it may strongly depend on $T_{\rm RH}$.
In this work, we scan $T_{\rm RH}$ from the $\sim$ GeV scale up to several tens of TeV.

At sufficiently low reheating temperatures $T_{\text{RH}}$, the cutoff scale $\Lambda$, which controls the interaction strength between the dark sector and the SM, must decrease by several orders of magnitude in order to reproduce the observed relic abundance.
This, in turn, enhances the decay width of $\phi$ and leads to lifetimes that can fall within the range relevant for laboratory-based and collider experiments.
\begin{figure}
    \centering
    \includegraphics[width=0.55\linewidth]{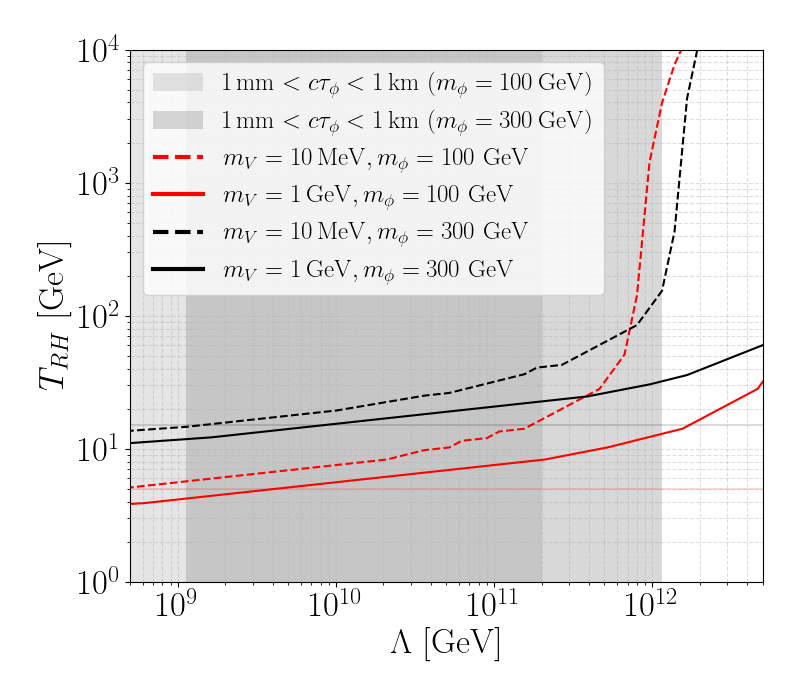}
    \caption{Contours of the correct relic abundance for four benchmark points of $(m_V, m_\phi)$, assuming $\lambda_{HS}=1$. The gray regions indicate the parameter regions relevant for collider phenomenology, where $\phi$ behaves as an LLP. The red contours should be compared with the light-gray regions, while the black contours should be compared with the dark-gray regions. The horizontal lines correspond to the freeze-out temperature $T_f$ of $\phi$ in each case.}
\label{fig:relic_solution}
\end{figure}
In figure~\ref{fig:relic_solution}, we take several benchmark points to illustrate this behavior.
The red and black contours correspond to the observed relic abundance for $m_\phi = 100$ GeV and 300 GeV, respectively, fixing $\lambda_{HS}=1$, and for different values of $m_V$ (dashed for $m_V=10$~MeV and solid for $m_V=1$~GeV).
When $T_\text{RH}$ becomes sufficiently small, the Boltzmann suppression becomes relevant, leading to smaller values of $\Lambda$, or equivalently to larger effective couplings $c_{5}/\Lambda$, where we fix the Wilson coefficient to $c_{5}=1$.
The behavior is qualitatively similar in all four cases.
However, the strong variation with $T_\text{RH}$ is more pronounced for smaller values of $m_V$ at fixed $m_\phi$, as shown in the figure.
For heavier $m_V$, this behavior is shifted towards larger values of $\Lambda$.
Furthermore, we highlight with gray the parameter regions where $\phi$ behaves as an LLP potentially accessible at colliders.
\begin{figure}
    \centering
    \includegraphics[width=1\linewidth]{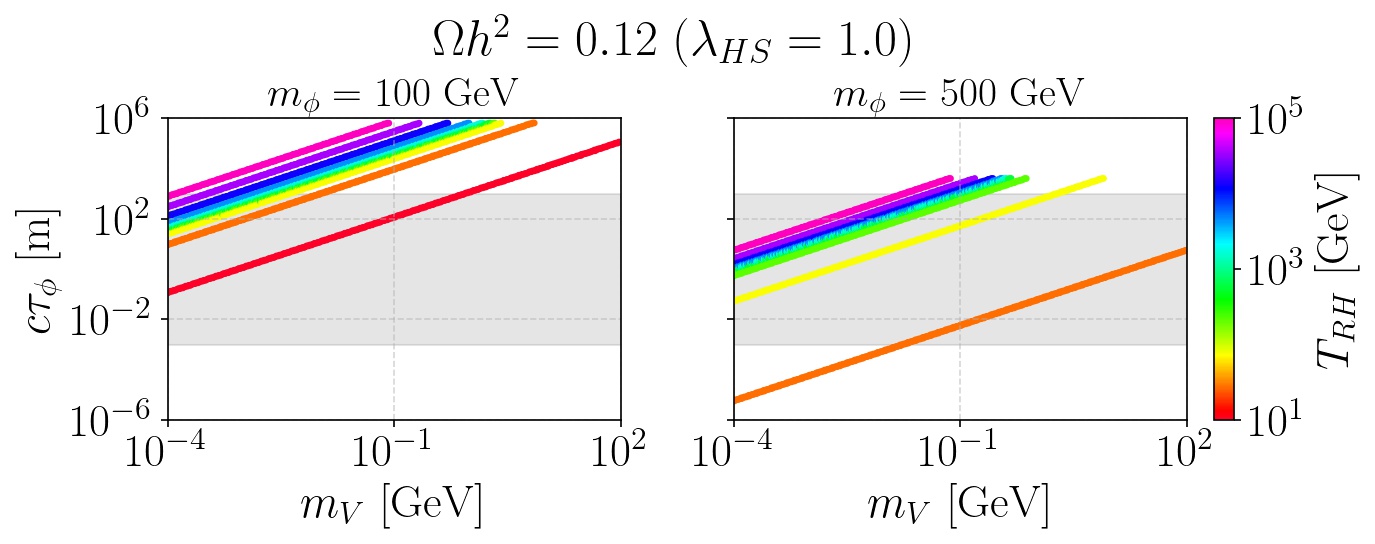}
    \caption{Parameter space that saturate the relic density in the plane ($m_V,c\tau_\phi$) for different values of $m_\phi$. We highlight the LLP target region where $c\tau_\phi \in [10^{-3},10^3]$ m. All the points fulfill $T_{\rm RH}>T_f$.  }
    \label{fig:relic_ctau_mass}
\end{figure}

In conclusion, relatively low values of $T_\text{RH}$ are generally required in the parameter region of our interest, not only to reproduce the observed relic abundance via the FI mechanism, but also to lower $\Lambda$ to values around $10^9 -10^{12}$ GeV.
This, in turn, leads to $c\tau_\phi$ values in the range relevant for searches for displaced $\phi$ decays at main LHC  detectors and far detectors.

For definitiveness, to study the collider phenomenology in Sec.~\ref{sec:collider} we set $m_V = 10$ MeV, as this choice opens up a wider range of possibilities for $T_{\rm RH}$ within the region of our interest while having $\phi$ as a long-lived state, spanning both low and high values (see figure~\ref{fig:relic_ctau_mass}). In this context, we do not restrict our analysis to low reheating temperatures, but instead show that the LHC and FCC-hh can also probe scenarios with high $T_{\rm RH}$.

\subsection{(In)direct detection and cosmological bounds}
As $\Lambda$ lies at high scales, the dark photon $V$ interacts only feebly with ordinary matter, rendering direct detection experiments largely insensitive to this scenario.
In addition, the direct detection cross-section is loop-suppressed, as no tree-level couplings between $V$ and SN particles are present.\footnote{For similar conclusions in the \textit{co-scattering} regime, see ref.~\cite{DiazSaez:2024dzx}.}
Likewise, indirect detection signals from dark matter annihilation are expected to be highly suppressed as a result of the smallness of the relevant couplings.
Consequently, the most promising probes of this framework are collider searches for LLPs, together with cosmological observations sensitive to the dark matter relic abundance.

On the other hand, once $\phi$ enters in chemical equilibrium with the SM, it decouples while being non-relativistic and subsequently decays, injecting additional dark photons.
The velocity of these dark photons and the epoch at which $\phi$ decays depend on the mass splitting between the dark states and on the scale $\Lambda$, respectively.
For the parameter region considered here, any possible dark photon radiation during the BBN epoch is completely negligible, mainly for two reasons.
First, although the dark photons produced in the decay of $\phi$ are initially highly relativistic, at the time of BBN, most of the dark photons arrive being non-relativistic. Second, the population of the dark photons produced in the out-of-equilibrium decay of $\phi$ is strongly suppressed, since $\phi$ never dominates the energy density of the Universe (see figure~\ref{fig:sw}).
The corresponding energy density therefore satisfies
\begin{equation}
\frac{\rho_V^{\rm decay}}{\rho_{\rm rad}}
\sim
\frac{n_\phi E_{V}}{\rho_{\rm rad}}
\ll 1 ,
\end{equation}
implying a negligible contribution to $\Delta N_{\rm eff}$ at the time of BBN.
Since no significant extra radiation is present at BBN temperatures, the impact at later cosmological epochs is even further suppressed. For further details, see Appendix~\ref{App}.

\section{Collider phenomenology}\label{sec:collider}

We study the collider signatures of our long-lived scalar, residing in a parameter region consistent with the assumption of vector DM.
The scalar can have lifetimes spanning from picosecond to microsecond (which corresponds to $c\tau_\phi$ values from $\sim 10^{-4}$ to $10^2$ m; see figure~\ref{fig:relic_solution}), allowing it to be tested by multiple searches both at the LHC and FCC-hh main detectors and proposed far detectors.
The scalar $\phi$ can decay to either $\gamma V$ or $Z V$, with the corresponding partial decay widths given in equation~\eqref{eqn:Gamma_phi2gammaV} and equation~\eqref{eqn:Gamma_phi2ZV}.
The total decay width and the proper decay length of $\phi$ can be found in equation~\eqref{eqn:Gamma_phi} and equation~\eqref{ctau}, respectively. In this section, we set $m_{V}=10^{-2}$ GeV, as it maximizes the range of reheating temperature that can be probed with LLP searches (see figure \ref{fig:relic_ctau_mass}).

\begin{figure}[h]
    \centering
    \includegraphics[width=0.5\linewidth]{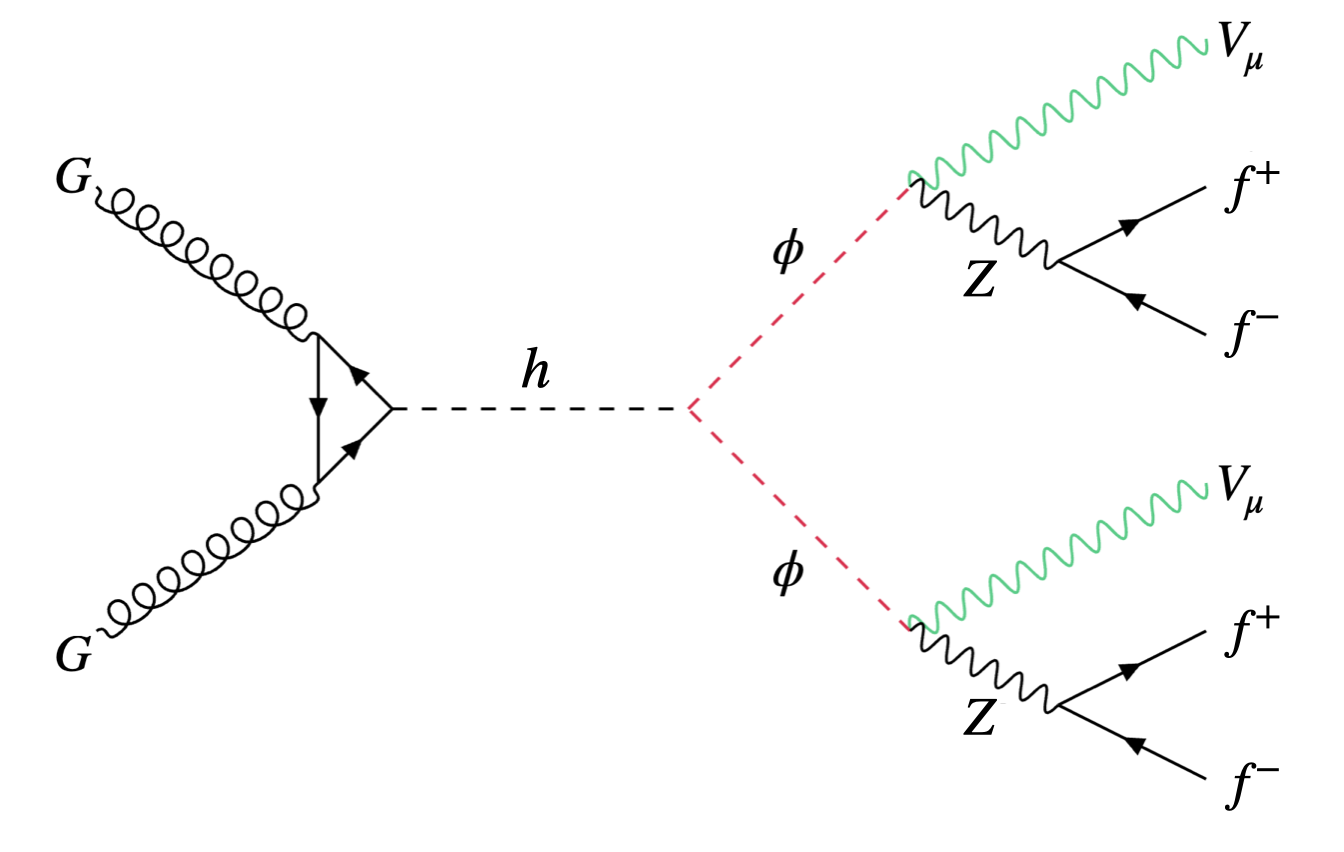}
    \caption{Dark photon DM production at proton-proton colliders. The $\phi$-production is mediated by the $\lambda_{HS}$ coupling while its decay proceeds via $\Lambda$. Note that, in the kinematic region of our interest, the Higgs boson is off-shell.}
    \label{fig:process}
\end{figure}

As we focus on the decay $\phi\to ZV$, the $Z$-boson decays into visible and electrically charged fermions, leading to displaced signatures at both the LHC main detectors and far detectors such as MATHUSLA~\cite{Chou:2016lxi,Curtin:2018mvb,MATHUSLA:2018bqv,MATHUSLA:2020uve,MATHUSLA:2025zyt} and ANUBIS~\cite{Bauer:2019vqk,ANUBIS:2025sgg}.
The latter usually consist of tracking detectors without calorimeters, as opposed to general-purpose detectors such as \texttt{ATLAS} or \texttt{CMS}.
The complete process we consider is sketched in figure~\ref{fig:process}. 
In order to compute this process, we implemented the model using \texttt{Feynrules}~2.3.49~\cite{Alloul:2013bka} interfaced with \texttt{Feynarts}~3.12~\cite{Hahn:2000kx} and \texttt{NLOCT}~1.02~\cite{Degrande:2014vpa} to generate UFO files compatible with \texttt{MadLoop}~3.6~\cite{Hirschi:2011pa}.
With these files, we compute the process beyond tree level without requiring effective couplings between gluons and the Higgs boson.
 
Furthermore, we study the search prospect at FCC-hh, considering its cross-sections and integrated luminosities are larger than those at the LHC.

\subsection{LHC prospects with main and far detectors}

We consider several LLP search strategies at the LHC main detectors and proposed far detectors, that provide complementary sensitivity reach to our model parameters.

\vspace{1cm}

{\bf{LHC main detectors.}}
We focus on two LHC search strategies at $\sqrt{s}=13$~TeV.
We first recast an \texttt{ATLAS} ``displaced vertex (DV)+MET" search~\cite{ATLAS:2017tny}.
This search was first recast by some of us in ref.~\cite{Proceedings:2018het}; see also ref.~\cite{Desai:2021jsa}, where some of us recast the same search in the framework of \texttt{CheckMATE}~\cite{Drees:2013wra,Dercks:2016npn}.
It was also recently implemented in ref.~\cite{Arbelaez:2024lcr} to constrain a model of scotogenic dark matter.
The main observable of the DV+MET search is a high-mass (larger than 10 GeV) and high-track-multiplicity (larger than 5 displaced tracks) secondary vertex, reconstructed inside the \texttt{ATLAS} inner tracker (with displacements between 4 and 300 mm).
The search also requires a trigger cut on missing transverse momenta (i.e.~MET or $p_T^{\text{miss}}$) of 200 GeV.

In principle, it is also possible to probe the scenario with a search strategy for time-delayed leptons~\cite{Liu:2018wte}, which uses the upcoming high-precision timing detectors to be installed at the \texttt{ATLAS} and \texttt{CMS} detectors.
Such a ``time-delay'' search is complementary to direct DV searches, as it proposes to measure time delays as opposed to (displaced) information from the trackers, with the timing detectors.
However, owing to the relatively tiny decay branching ratios of the scalar $\phi$ into a pair of charged leptons and DM, the time-delay search is found to have no sensitivity to our scenario.
Therefore, we will focus on the DV+MET search below to estimate the reach at LHC main detectors.

\begin{figure}[t!]
    \centering
    \includegraphics[width=0.6\linewidth]{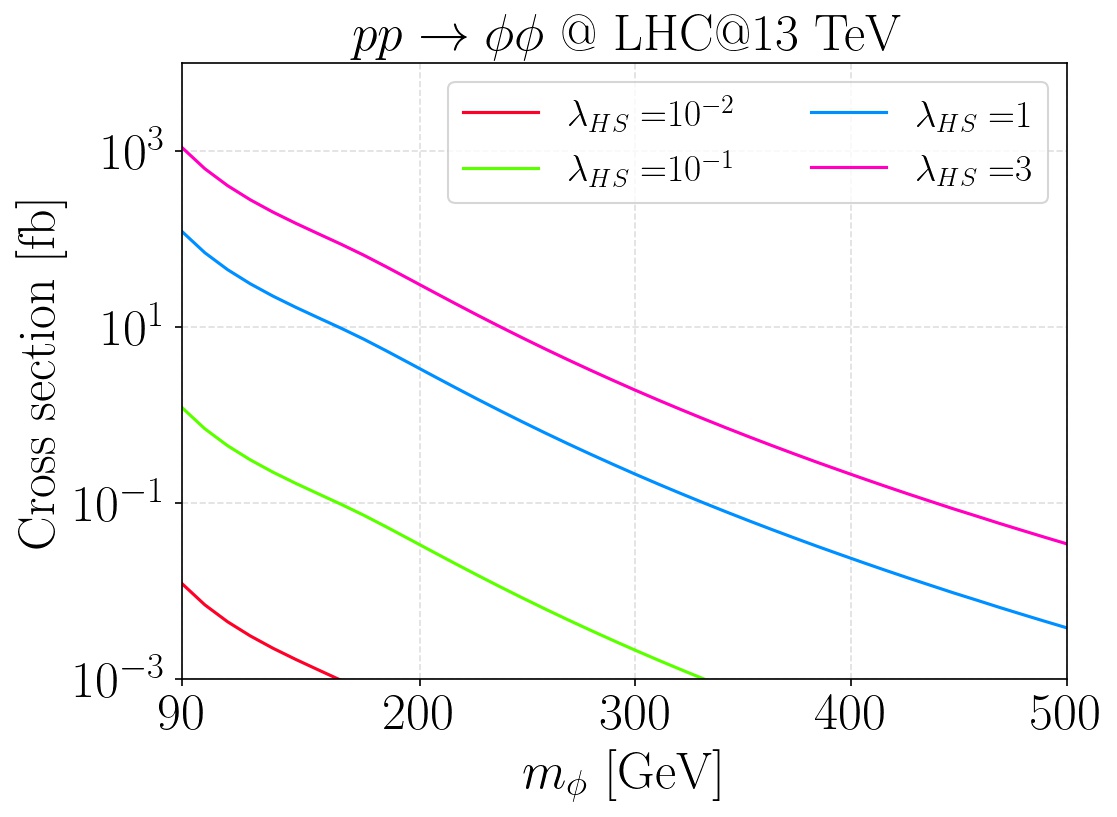}
    \caption{Parton-level production cross-sections at the LHC with a center-of-mass energy of 13~TeV. }
    \label{fig:xsection}
\end{figure}

We show in figure~\ref{fig:xsection} parton-level production cross-sections at the LHC, computed with {\texttt{MadGraph5\_aMC@NLO}}~\cite{Alwall:2011uj,Alwall:2014hca} version 3.6.2 for several fixed values of the Higgs-portal coupling $\lambda_{HS}$, as functions of the scalar mass $m_\phi$.
Generated events are then interfaced to \texttt{Pythia8}~\cite{Sjostrand:2014zea} version 8.308  for showering and hadronization.
The long-lived scalar is also set to decay in \texttt{Pythia8}, when interfacing with a customized code that reconstructs DVs and missing transverse momenta (see refs.~\cite{Proceedings:2018het,Arbelaez:2024lcr} for more details).

As the original \texttt{ATLAS} DV+MET search has a too high threshold of 200~GeV on MET, this leads to a large reduction of signal events and thus no sensitivities to our model scenario.
We therefore follow ref.~\cite{Arbelaez:2024lcr} and propose an optimized search lowering this trigger cut to 50~GeV.\footnote{This could be justified by requiring additional triggers on displaced activity~\cite{Alimena:2021mdu}, as also suggested in ref.~\cite{Cottin:2022nwp} when optimizing a different search for heavy neutral leptons.}
With this search, zero backgrounds are expected and we thus will take 3 signal events as the exclusion limit at 95\% confidence level (CL).

\begin{figure}[h!]
    \centering
    \includegraphics[width=\linewidth]{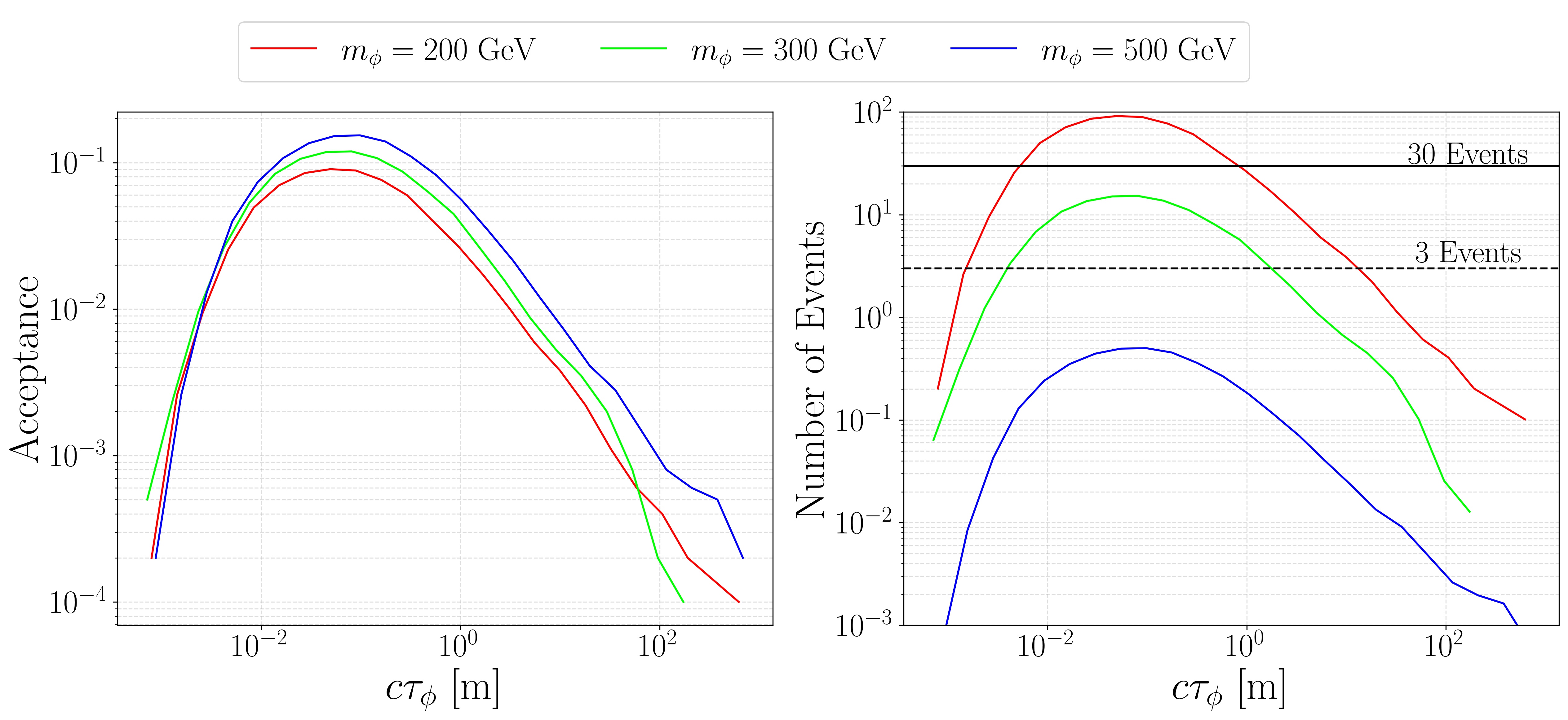}
    \caption{Acceptance and number of expected events with the \texttt{ATLAS} DV+MET search, for a trigger cut of MET$>50$~GeV, as function of the lifetime and the mass of $\phi$. We have set $\lambda_{HS}=3$.}
    \label{fig:DVMET}
\end{figure}

In figure~\ref{fig:DVMET} we show plots of the search acceptance and the predicted signal-event numbersfor a trigger cut of MET $>50$~GeV and $\lambda_{HS}=3$, as functions of $c\tau_\phi$, for $m_\phi=200$~GeV (red), 300~GeV (green), and 500~GeV (blue).
These acceptances include parametrized event-level and vertex-level efficiencies provided by ATLAS~\cite{ATLAS:2017tny}, as functions of LLP decay distance, LLP mass, number of tracks associated with the DV, and MET.
For decays beyond the ATLAS calorimeter (which happen more often at high $c\tau_\phi$) the efficiency values provided remain mostly constant, independently of the value of MET.

The mountain-shaped curves reflect that fact that for too small (large) decay lengths, the scalar $\phi$ decays too promptly (far), outside the fiducial volume of the search.

{\bf{LHC far detectors.}} In order to study the projected sensitivities to our model scenario at the proposed LHC far detectors, we employ the tool \texttt{Displaced Decay Counter} (\texttt{DDC})~\cite{Domingo:2023dew}.
\texttt{DDC} reads the \texttt{MadGraph5}-generated LHE event samples as input and computes the detector acceptance to the long-lived $\phi$ and thus the expected number of signal events at the far detectors, with the following formula,
\begin{equation}
    \mathcal{N} =\sigma \cdot \mathcal{L}\cdot \langle P_{\text{fiducial}}\rangle \cdot \text{Br}(\phi \to V f^+ f^-)\cdot \epsilon,
\end{equation}
where $\langle P_{\text{fiducial}}\rangle$ is the average probability of the long-lived scalars to decay inside the fiducial volume of a far detector (acceptance), and $\epsilon$ denotes the detector reconstruction efficiencies, which is assumed to be $100\%$.
The acceptance is computed by taking into account the LLP mass and proper lifetime, LLP kinematics, as well as the position and geometries of the far detectors, and exploiting exponential decay distribution functions.
We restrict this part of the study to the proposed \texttt{MATHUSLA} and \texttt{ANUBIS} experiments, considering their relative advantages in detecting LLPs with masses from tens to hundreds of GeV.
We take into account the recent design modifications in both experiments~\cite{MATHUSLA:2025eth,ANUBIS:2025sgg}.
In particular, the size reduction of the \texttt{MATHUSLA} design reduces the acceptance by about one order of magnitude.
On the other hand, the latest design of \texttt{ANUBIS} (now on the cavern ceiling instead of inside one service shaft) enhances the acceptance despite higher SM background levels.
According to ref.~\cite{ANUBIS:2025sgg}, a background of $182 \pm 12$ events are expected at $\mathcal{L}=3~$ab$^{-1}$.
The original and most updated designs have been implemented in \texttt{DDC}~\cite{Wang:2025esc}.

\begin{figure}
    \centering
    \includegraphics[width=0.5\linewidth]{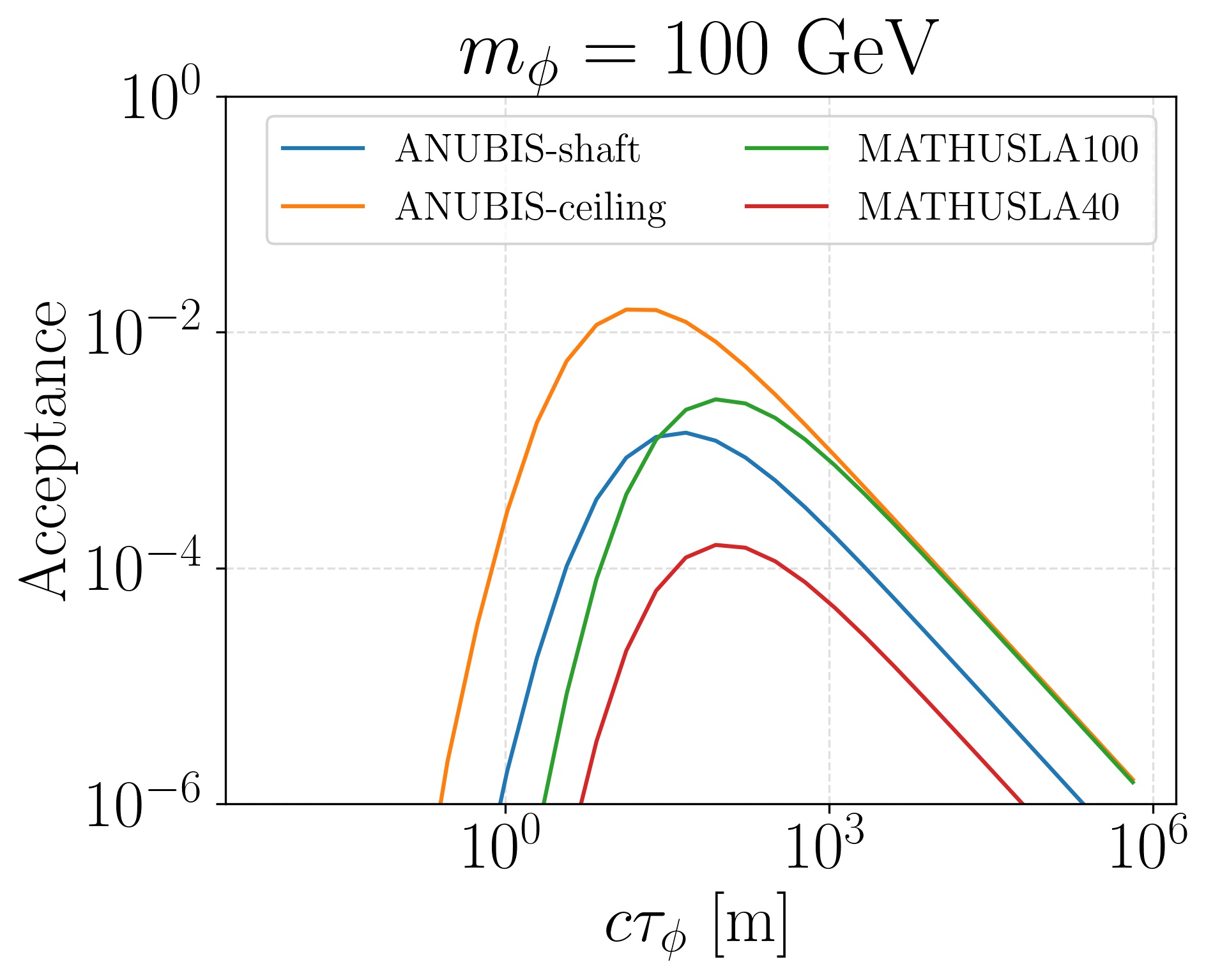}
    \caption{Acceptances of the various far detector proposals at the LHC, as functions of $c\tau_\phi$.  }
    \label{fig:acceptance_far_detector}
\end{figure}

Figure~\ref{fig:acceptance_far_detector} shows the acceptances of \texttt{ANUBIS-shaft}, \texttt{ANUBIS-ceiling}, \texttt{MATHUSLA100}, and \texttt{MATHUSLA40}, as functions of $c\tau_\phi$ for a fixed mass of 100~GeV.
Note that the choice of $\lambda_{HS}$ affects the production cross-sections, but not the acceptance.
\texttt{ANUBIS-shaft} and \texttt{MATHUSLA100} are expected to have comparable acceptances while \texttt{ANUBIS-ceiling} and \texttt{MATHUSLA40} should have the largest and smallest probabilities to have the long-lived $\phi$ decays inside them.

\subsection{FCC-hh prospects with main and far detectors}

The signal-event rates are strongly suppressed by the $Vf^+f^-$-channel branching fractions of $\phi$ and therefore the LHC predictions are testable only for a limited parameter region.
We proceed thus to study the sensitivities expected at the FCC-hh~\cite{Benedikt:2022kan}, considering a center-of-mass energy of $\sqrt{s}=100$~TeV and an integrated luminosity of $\mathcal{L}=20$~ab$^{-1}$.
Under these experimental conditions, we expect substantial enhancement in the signal-event yields, allowing to probe wider regions in model parameter space.

The FCC-hh main detector design~\cite{FCC:2018vvp} includes a \texttt{Central Tracker} (\texttt{CT}) and a \texttt{Forward Tracker} (\texttt{FT}).
For both the \texttt{CT} and \texttt{FT}, we employ a simple geometrical event-counting approach as in the case of the LHC far detectors, without a dedicated search strategy (especially for the \texttt{CT}), assuming vanishing background levels.
This simplified treatment should be perceived as a tentative approach as the FCC-hh is still in the design state and its operation would only take place decades from now.

Following ref.~\cite{FCC:2018vvp} (see ref.~\cite{Bernal:2025qkj} for a simpler modeling outside the \texttt{DDC}), we model the \texttt{CT} as an annular cylinder, with a length of 5~meters and an inner (outer) radius of 20~mm (1.7~m).
The FCC-hh \texttt{FT} is to be placed in the forward region along the beam ($z$) axis.
It has a cylindrical shape of a length of 6~m.
Its front end is 10~meters away from the interaction point (IP), and its inner (outer) radius is 20~mm (1.6~m).

Besides the main detectors, we consider far detector proposals at the FCC-hh, called \texttt{DELIGHT} and \texttt{FOREHUNT}~\cite{Bhattacherjee:2021rml,Bhattacherjee:2023plj}.
\texttt{DELIGHT} is a transverse detector symmetrical along the $z$-direction but with a large distance from the IP in the transverse plane.
Ref.~\cite{Bhattacherjee:2021rml} proposed three configurations of \texttt{DELIGHT} and we focus on the smallest one here, \texttt{DELIGHT-A}.
\texttt{DELIGHT-A} covers $z$ in $[-50~\text{m}, 50~\text{m}]$, $x$ in $[25~\text{m}, 50~\text{m}]$, and $y$ in $[0~\text{m}, 100~\text{m}]$.

\texttt{FOREHUNT} is a forward cylindrical detector proposed with three possible configurations in ref.~\cite{Bhattacherjee:2023plj}.
We study the \texttt{FOREHUNT-C} setup which has a radius of 5~m, a length of 50~m, and a nearest distance of 50~m in front of the IP.

\begin{figure}[htbp]
    \centering
    \includegraphics[width=0.5\linewidth]{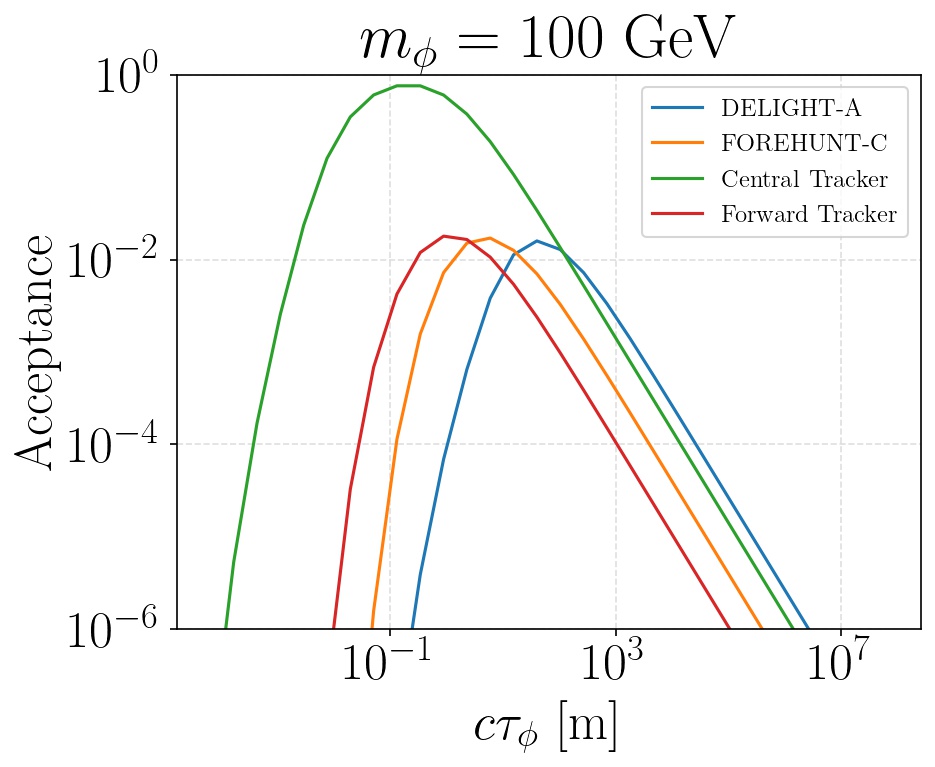}
    \caption{Acceptances as functions of $c\tau_\phi$ for the different detectors at the FCC-hh including both main and far detectors.}
    \label{fig:acceptance_fcchh}
\end{figure}

All these detector setups have been implemented in \texttt{DDC} which is then used for evaluating the signal acceptance and yields. The acceptance for these detectors can be seen in figure~\ref{fig:acceptance_fcchh}, where we notice that the \texttt{Central Tracker} outperforms the other proposals in general.

\section{Numerical results}\label{sec:results}

In this section we present and discuss the collider sensitivities, as well as their interplay with the reheating temperature.
Given signal events observed in one or more of these collider experiments, we could explain the DM relic density by fitting the reheating temperature.
In this section, we define two types of contours. Dashed (solid) lines correspond to at least 3 (30) signal events for all experiments, except for \texttt{ANUBIS-ceiling}, where the corresponding thresholds are 14 (81) events.

\begin{figure}
    \centering
    \includegraphics[width=0.6\linewidth]{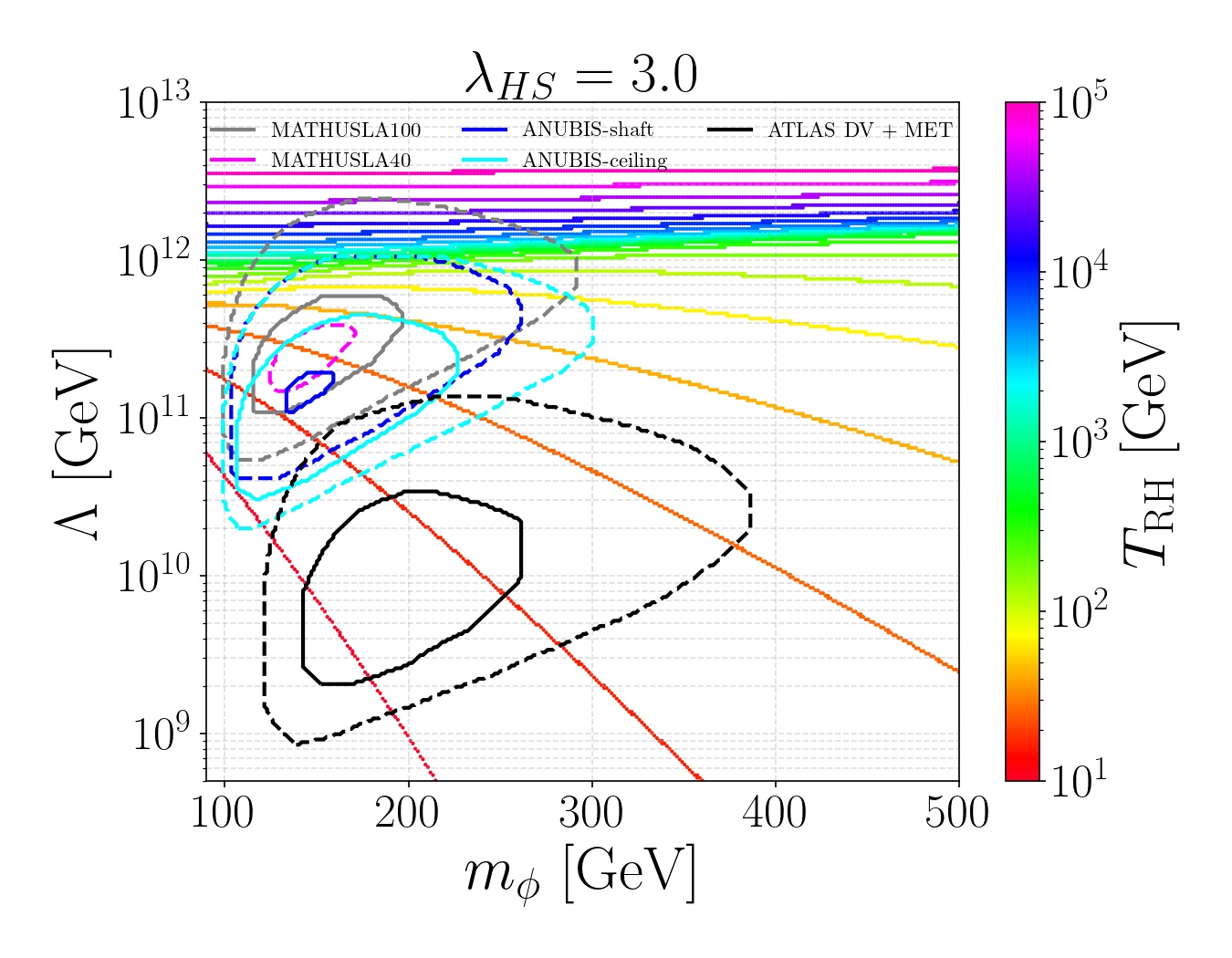}
    \caption{Sensitivity of LHC searches considering both main and far detectors,
    with the reheating temperature required to fullfil the DM  relic density. The contours were obtained at $\mathcal{L}=3$~ab$^{-1}$. Dashed lines represent optimistic predictions, while solid lines are more conservative ones. }
    \label{sensitivity_lhc}
\end{figure}

We now discuss this interplay expected at the LHC and the FCC-hh separately.
As shown in figure~\ref{sensitivity_lhc}, the HL-LHC with an integrated luminosity of 3~ab$^{-1}$ can probe the mass of the mediator in the range of $90\lesssim  m_\phi\lesssim 400$~GeV. 
In addition, far detectors can test larger values of $T_{\text{RH}}$, while the DV strategy can test the low reheating scenario.
The combined results allow to set bounds on $T_{\text{RH}}$ as
\begin{equation}
    10~\text{GeV} \leq T_{\text{RH}}^{(\text{LHC})}\leq 10^3~\text{GeV}.
\end{equation}

An important feature is that there is some overlap in sensitivity between the main detector and the far detectors, highlighting their complementarity in probing different regions of parameter space.
Another relevant aspect is that the downsize of the \texttt{MATHUSLA} design reduces significantly the sensitivity; however, there is an interesting complementarity between far detectors: signals in both \texttt{MATHUSLA} and \texttt{ANUBIS} could help constrain the reheating temperature even more.

Finally, contours close up only in the region where $\lambda_{HS}=3$. For lower values of this coupling, the event yields are too small and we are not able to define conservative contours. This rather strict choice of the Higgs-portal coupling is what motivate us to consider also the prospects at the FCC.

\begin{figure}[t]
    \centering
    \includegraphics[width=\linewidth]{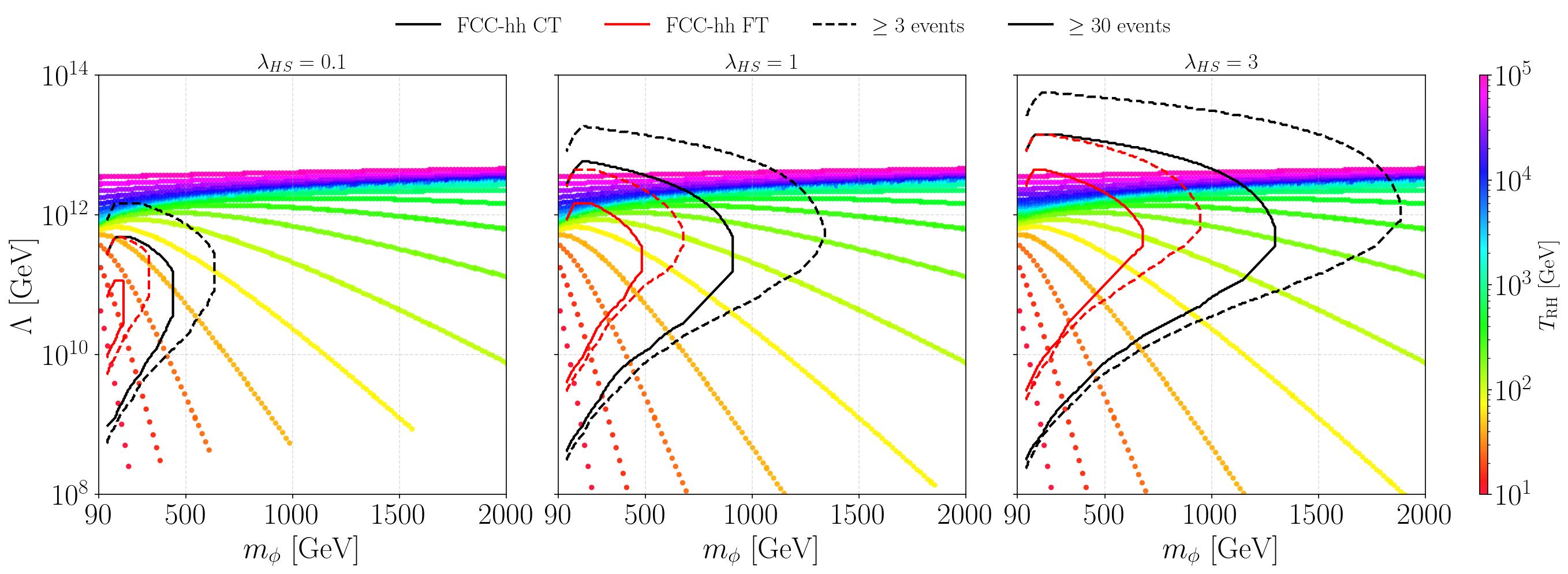}
    \caption{Sensitivity of the main FCC-hh detectors accompanied with the reheating temperature required to fulfill the DM relic density. The contours are for an integrated luminosity of $\mathcal{L}=20$~ab$^{-1}$.}
    \label{sensitivity_fcc_main}
\end{figure}

\begin{figure}[t]
    \centering
    \includegraphics[width=\linewidth]{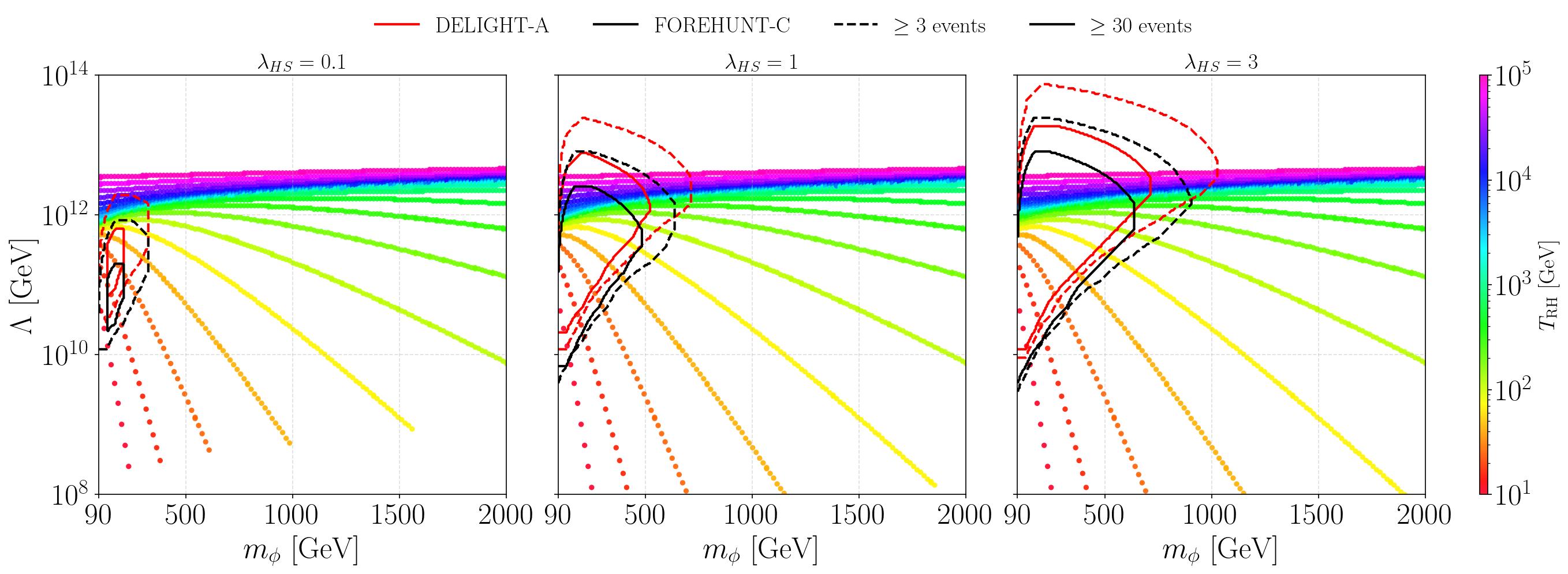}
    \caption{Sensitivity of DELIGHT and FOREHUNT accompanied with the reheating temperature required to fulfill the DM relic density. The contours correspond to an integrated luminosity of $\mathcal{L}=20$~ab$^{-1}$.}
    \label{sensitivity_fcc_far}
\end{figure}

For the FCC-hh, the sensitivity results are shown in figures~\ref{sensitivity_fcc_main} and~\ref{sensitivity_fcc_far}, for the main detector and the proposed far detectors, respectively.
For this collider, we can probe mediator masses up to the TeV scale, and for the coupling $\lambda_{HS}$ even as small as $0.1$, we still expect a significant number of events.
The main \texttt{Central Tracker} is shown to have stronger sensitivity reach, covering the sensitivity curves of the forward tracker completely.
Thus, within the parameter regions covered by the FCC-hh forward tracker, it is possible to observe events at both the FCC-hh central and forward trackers.
Also, we are able to probe a wider span of $T_{\text{RH}}$ than with the LHC experiments
\begin{equation}
    10~\text{GeV} \leq T_{\text{RH}}^{(\text{FCC})}\leq 10^5~\text{GeV}.
\end{equation}

In addition, as shown in figure~\ref{sensitivity_fcc_far}, mainly owing to the large luminosity at the FCC-hh and the relatively close distance of the propose far detectors from the IP, the sensitivities of the far detectors such as DELIGHT-A and FOREHUNT-C are only mildly weaker than those at the main trackers.

\section{Conclusions}\label{sec:conclusions}
We have explored an extension of the singlet scalar Higgs-portal model, introducing a dark sector with a vector field $V_\mu$ as the dark matter candidate and a real scalar $\phi$ that decays via a higher-dimensional operator, generically referred to as the dark-axion portal.
This portal enables the (displaced) decay channel $\phi \to Z + V$, followed by $Z \to f^+ f^-$, leading to distinctive LLP signatures.
In this framework, $\phi$ serves as a crucial link between the visible and dark sectors.

We have studied dark matter production in the region of parameter space where $\phi$ behaves as an LLP.
We find that freeze-in is the dominant production mechanism for dark matter, both at high and low reheating temperatures.
The Super-WIMP contribution to the relic abundance turns out to be subleading and consistent with BBN bounds. Furthermore, we find that both direct and indirect detection prospects are suppressed in the region of parameter space under consideration.

We have also investigated the collider phenomenology of the model, focusing on the prospects of observing displaced decays of $\phi$ at the LHC, considering both main and far detectors.
In particular, we have studied the sensitivity for MATHUSLA and ANUBIS.
We find strong complementarity between these two classes of detectors, and sizable event rates for a well defined region of the parameter space.

In addition, we have estimated the projected sensitivities at the FCC-hh, and our results show that larger regions of the parameter space can be probed. For the FCC, we notice that most of the parameter space is covered by the \texttt{Central Tracker} of the main detector, and the far detectors could provide a validation of our model if a signal is observed.

Our results demonstrate that LLP searches can indirectly probe the reheating temperature, as displaced signatures stemming from a long-lived mediator to the dark sector could place constraints across 5 orders of magnitude.

\section*{Acknowledgments}
GC would like to thank the School of Physics at the Hefei University of Technology for hospitality offered while working
on this project. GC acknowledges support from ANID FONDECYT grant No. 1250135. BDS was partially funded by ANID FONDECYT Iniciación No.~1125181. PA acknowledges support from ANID BECAS/DOCTORADO NACIONAL 21250555. PA, GC and BDS also acknowledge support from ANID – Millennium Science Initiative Program ICN2019\_044 (Millennium Institute for Subatomic Physics at the High Energy Frontier, SAPHIR). ZSW and YZ acknowledge support from the National Natural Science Foundation of China under grants No.~12475106 and No.~12505120, and the Fundamental Research Funds for the Central Universities under Grant No.~JZ2025HGTG0252.

\appendix

\section{Extra radiation at BBN and the warmness of the DM} \label{App}
\subsection{Decay temperature}

Recall that we consider $m_\phi > m_Z \gg m_V$, such that the total decay width of $\phi$ in equation~\eqref{eqn:Gamma_phi} reduces to
\begin{equation}
\Gamma_{\rm \phi}
=
\Gamma(\phi \to \gamma V)
+
\Gamma(\phi \to Z V)
\simeq
\frac{m_\phi^3}{8\pi \Lambda^2}. \label{tot_decay}
\end{equation}
The characteristic decay temperature $T_D$ can be estimated by equating the decay rate to the Hubble expansion rate $\Gamma_\phi \simeq H(T_D)$.
Solving the latter equation for a radiation dominated universe, and replacing the decay width in equation~\eqref{tot_decay}, the decay temperature can be written as
\begin{equation}
T_D \approx
\left(\frac{45}{4\pi^3 g_*}\right)^{1/4}
\sqrt{\frac{M_{\rm Pl}}{8\pi}}
\frac{m_\phi^{3/2}}{\Lambda}. \label{TD}
\end{equation}
If $T_D > T_f$, the decay rate is already much larger than the Hubble expansion at freeze-out, $\Gamma_{\phi} \gg H(T_f)$.
While inverse processes maintain equilibrium at early times, after chemical decoupling the relic $\phi$ abundance is rapidly depleted by decays.
In this regime, decays become effectively instantaneous after freeze-out, so that $T_f$ sets the relevant timescale for the depletion of the $\phi$ population.

\subsection{Momentum redshift}

Once $\phi$, being out-of-equilibrium, decays into $\gamma,Z$ plus $V$, the dark photon has its three-momentum magnitude given by
\begin{equation}
p_V(T_D) \approx \frac{m_\phi}{2}\,.
\end{equation}
Assuming adiabatic expansion, the three-momentum redshifts as
\begin{equation}
p_V(T) = p_V(T_D)\,\frac{T}{T_D}
= \frac{m_\phi}{2}\,\frac{T}{T_D}\,.
\end{equation}
The energy of the vector particle is therefore
\begin{equation}
E_V(T)
=
\sqrt{p_V(T)^2 + m_V^2}
=
\sqrt{
\left(\frac{m_\phi}{2}\frac{T}{T_D}\right)^2 + m_V^2
}\,.
\end{equation}
The transition to the non-relativistic regime occurs when $p_V(T_{\rm NR}) \simeq m_V$
which gives
\begin{equation}
T_{\rm NR}
=
T_D \,\frac{2 m_V}{m_\phi}\,.
\end{equation}
For illustration, we consider representative values
\begin{equation}
m_\phi = 100\,{\rm GeV}, 
\qquad
m_V = \{0.01,\; 1\},{\rm GeV},
\qquad
\Lambda = \{10^{9},\;10^{10},\;10^{11},\;10^{12}\}\,{\rm GeV}.
\end{equation}

\begin{figure}[h]
    \centering
    \includegraphics[width=0.45\linewidth]{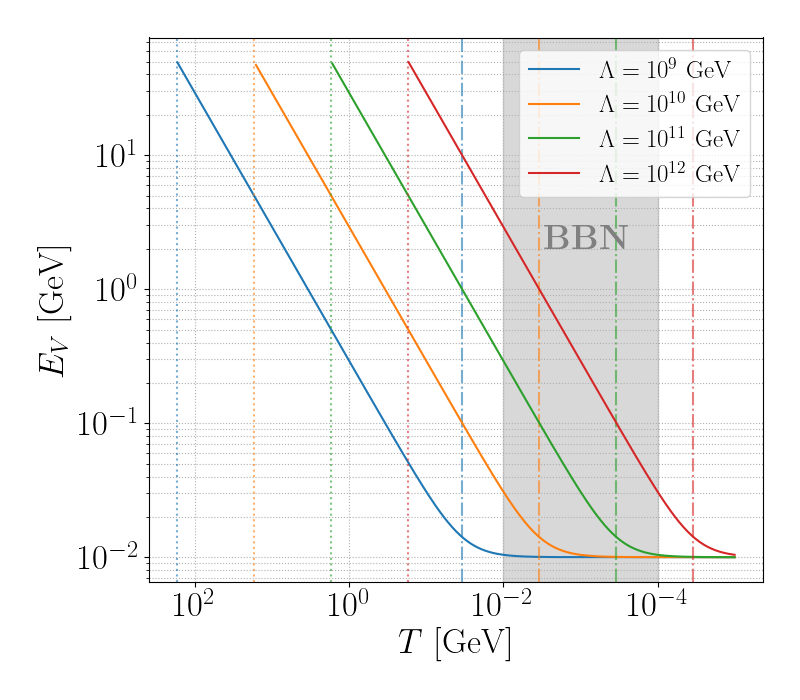}
    \includegraphics[width=0.45\linewidth]{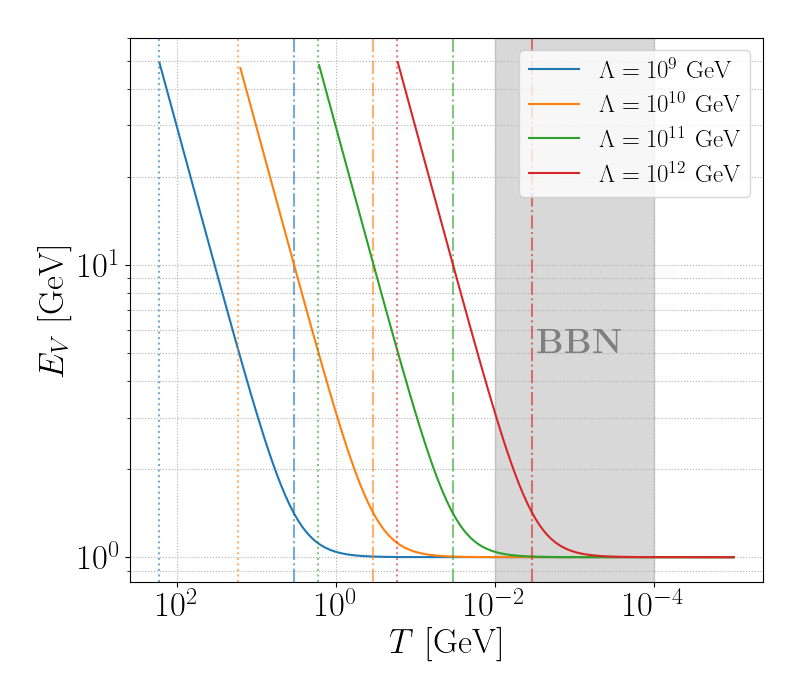}
    \caption{Dark photon energy from $\phi$ decays as a function of temperature for $m_V$ = 10~MeV (the left panel) and 1~GeV (the right panel), for different values of $\Lambda$. In both panels we consider $m_\phi = 100$ GeV. The gray band indicates the period of BBN, and the dotted and dotted-dashed vertical lines indicate $T_D$ and $T_{\rm NR}$, respectively.}
    \label{fig:BBN}
\end{figure}

The corresponding behavior of $E_V(T)$ is shown in figure~\ref{fig:BBN}, where we also indicate $T_D$ and $T_{\rm NR}$ for each parameter choice.
For $T \gg T_{\rm NR}$, the dark photon $V$ is relativistic and
\begin{equation}
E_V(T) \simeq p_V(T) \propto T\,,
\end{equation}
while for $T \ll T_{\rm NR}$ it becomes non-relativistic and
\begin{equation}
E_V(T) \simeq m_V\,.
\end{equation}

This transition is governed by the hierarchy among $m_\phi$, $m_V$, and the decay temperature $T_D$.
In the left panel of Fig.~\ref{fig:BBN}, we observe that sizable values of $\Lambda$ lead to relativistic dark photons during BBN, since $T_D \propto \Lambda^{-1}$ (see equation~\eqref{TD}).
For the case of a larger $m_V$ (the right panel of Fig.~\ref{fig:BBN}), $E_V$ decreases more rapidly, thereby reducing the impact of relativistic dark photons during BBN compared to the lighter $V$ case, and thus alleviating the contribution to extra radiation. Finally, for a heavier $\phi$, this effect would be further mitigated, shifting the dot-dashed curves to the left in each panel.

\subsection{Dark photon population from $\phi$ decays and BBN constraints}

From figure~\ref{fig:BBN}~(left) we observe that for high $\Lambda$ values relativistic dark photons can be present during BBN.
The amount of dark radiation is crucial to assess its impact on BBN. Since the dark photon population is produced out of equilibrium from $\phi$ decays, it does not correspond to a thermal bath with a well-defined temperature.
Therefore, we estimate its contribution directly from its energy density.

As a simple and conservative estimate, we evaluate the energy density associated with the parent $\phi$ population:
\begin{equation}
\rho_{V}^{\rm decay}
\sim (n_\phi\, E_{V}) \,,
\end{equation}
with $n_\phi = Y_\phi\, s$, where $Y_\phi$ is the yield and $s$ is the entropy density. Taking representative values at $T = \mathcal{O}({\rm MeV})$ for relativistic dark photons~\cite{Arcadi:2017kky}
\begin{equation}
Y_\phi \sim 10^{-12}, 
\qquad
s \sim 10^{-9}\,{\rm GeV}^3,
\qquad
E_{V} \sim 1\,{\rm GeV},
\end{equation}
we obtain
\begin{equation}
\rho_{V}^{\rm decay}
\sim 10^{-21}\,{\rm GeV}^4.
\end{equation}
Assuming $g_* \approx 10$, the radiation energy density during BBN at $T \sim \mathcal{O}$(MeV) is
\begin{equation}
\rho_{\rm rad} \sim \frac{\pi^2}{30} g_* T^4 \sim 10^{-12}\,{\rm GeV}^4,
\end{equation}
allowing us to find that 
\begin{equation}
\frac{\rho_{\gamma'}^{\rm decay}}{\rho_{\rm rad}} \sim 10^{-9} \ll 1.
\end{equation}

Since the $\phi$ decays well before BBN, the resulting dark photon population subsequently redshifts as radiation and is further diluted by entropy injection in the SM plasma.
Therefore, the estimate given above should be regarded as conservative, as it neglects this additional suppression.

The contribution to the effective number of neutrino species can then be estimated as
\begin{equation}
\Delta N_{\rm eff} \sim \frac{\rho_{\gamma'}^{\rm decay}}{\rho_\nu} \ll 0.1,
\end{equation}
with $\rho_\nu \simeq (0.3 - 0.4)\,\rho_{\rm rad}$ at $T \sim \mathcal{O}({\rm MeV})$, well below the current BBN bounds~\cite{Fields:2019pfx}.

We conclude that, even in the most extreme region of parameter space (large $\Lambda$ and light $m_V$), the dark photon population produced in $\phi$ decays does not significantly affect BBN.
The cosmological dark matter abundance remains dominated by the freeze-in contribution.
This result is consistent with the conclusions of ref.~\cite{Arias:2025nub}, where a similar framework was analyzed.

\bibliographystyle{utphys}
\bibliography{References}
\addcontentsline{toc}{chapter}{\bibname}
\end{document}